\documentclass[aps,prl,twocolumn,showpacs]{revtex4}
\pdfoutput=1
\usepackage{times} 
\newcommand\eq[1] {(\ref{#1})}

\newcommand\fig[1] {\ref{fig:#1}}

\newcommand{\nonum}{\nonumber \\}
\newcommand{\beqa}{\begin{eqnarray}}
\newcommand{\eeqa}[1]{\label{#1}\end{eqnarray}}
\newcommand{\beq}{\begin{equation}}
\newcommand{\eeq}[1]{\label{#1}\end{equation}}

\newcommand{\Ga}{\alpha}

\newcommand{\Gd}{\delta}
\newcommand{\Ge}{\epsilon}

\newcommand{\Gg}{\gamma}

\newcommand{\Gx}{\xi}

\newcommand{\CC}{{\cal C}}

\newcommand{\CO}{{\cal O}}

\def\Bb{{\bf b}}

\def\Bd{{\bf d}}

\def\Bp{{\bf p}}

\def\Bu{{\bf u}}

\def\Bx{{\bf x}}

\def\Bz{{\bf z}}
\def\BA{{\bf A}}
\def\BB{{\bf B}}

\def \om {\omega}
\def \la {\lambda}

\def \RR {{\mathbb R}}
\def \CC {{\mathbb C}}

\def \ba {\begin{array}}
\def \ea {\end{array}}
\newtheorem {Thm} {Theorem} [section]

\newtheorem {Adef} [Thm] {Definition}

\newtheorem {Arem} [Thm] {Remark}

\newtheorem {Aexa} [Thm] {Example}

\newtheorem {Anot} [Thm] {Notation}

\def \refe #1.{(\ref{#1})}
\def \reff #1.{figure~\ref{#1}}
\def \refs #1.{section~\ref{#1}}
\def \refss #1.{subsection~\ref{#1}}
\def \refD #1.{Definition~\ref{#1}}
\def \refT #1.{Theorem~\ref{#1}}
\def \refL #1.{Lemma~\ref{#1}}
\def \refC #1.{Corollary~\ref{#1}}
\def \refP #1.{Proposition~\ref{#1}}
\def \refR #1.{Remark~\ref{#1}}
\def \refE #1.{Example~\ref{#1}}
\def \refN #1.{Notation~\ref{#1}}
\newcommand{\Mp}[1]{\left({#1}\right)}

\newcommand{\Bzero}{\mathbf{0}}

\newcommand{\figref}[1]{fig.~\ref{fig:#1}}

\usepackage{xcolor}
\usepackage{graphicx}
\usepackage{amssymb}
\usepackage{amsmath}
\usepackage{hyperref} 
\begin{document}
\title{Active Exterior Cloaking}
\author{Fernando Guevara Vasquez}
\author{Graeme W. Milton}
\author{Daniel Onofrei}
\affiliation{Department of Mathematics, University of Utah, Salt Lake City
UT 84112, USA}
\date{May 28 2009}
\begin{abstract}
A new method of cloaking is presented. For two-dimensional quasistatics
it is proven how a single active exterior cloaking device can be used to
shield an object from surrounding fields, yet produce very small
scattered fields. The problem is reduced to finding a polynomial which
is approximately one within one disk and zero within a second disk, and such
a polynomial is constructed. For the two-dimensional Helmholtz equation,
it is numerically shown that three active exterior devices placed around
the object suffice to produce very good cloaking.
\end{abstract}
\maketitle

Making a body truly invisible, in the sense of preventing it from
absorbing radiation or scattering radiation in any direction, has long
been regarded as the domain of science fiction rather than science, with
a few exceptions \cite{Dolin:1961:PCT, Kerker:1975:IB}.  This
perspective has changed in recent years due to fascinating advances in
our understanding of how electromagnetic fields can be manipulated.
Recent proposals for invisibility, as reviewed in
\cite{Alu:2008:PMC,Greenleaf:2009:CDE}, can be broadly placed in two
main groups: interior cloaking, where the cloaking device surrounds the
object to be cloaked, and exterior cloaking, where the cloaking region
is, surprisingly, outside the cloaking device.

Interior cloaking methods include plasmonic cloaking due to scattering
cancellation \cite{Alu:2005:ATP}, transformation based cloaking
\cite{Greenleaf:2003:ACC,Pendry:2006:CEM,Leonhardt:2006:OCM,Li:2008:HUC},
and active cloaking \cite{Miller:2007:PC}. 
Transformation based cloaking
has gained particular attention, being supported by experiment
\cite{Schurig:2006:MEC, Farhat:2008:BCA,Liu:2009:BGP,
Valentine:2009:OCD,Gabrielli:2009:COF} and rigorous mathematics
\cite{Greenleaf:2007:FWI,Kohn:2008:CCV,Kohn:2009:CCV}. While difficult
to exactly achieve, various approximate schemes make it more practical
\cite{Schurig:2006:MEC,Cai:2007:OCM,Cai:2007:NMC,Li:2008:HUC}.

Exterior cloaking methods include cloaking due to anomalous resonance
\cite{Milton:2006:CEA,Milton:2006:OPL,Nicorovici:2007:OCT,Milton:2008:SFG}
where polarizable dipoles, and polarizable line quadrupoles, and
clusters of arbitrarily many polarizable line dipoles in the vicinity of
a flat or cylindrical superlens
\cite{Veselago:1967:ESS,Nicorovici:1994:ODP,Pendry:2000:NRM} are
cloaked, but apparently not larger objects \cite{Bruno:2007:SCS}, and
cloaking due to complementary media \cite{Lai:2009:CMI} where an
``antiobject'' is embedded in a superlens, to create cancellation.

Miller \cite{Miller:2007:PC} found that active controls rather
than passive materials could be used to achieve interior cloaking.
Here we use active cloaking devices to
achieve exterior cloaking. In principle this could be done by
mimicking the effect of the cloaking device of \cite{Lai:2009:CMI}, but
our objective is to have an active cloaking device which does not
require one to know the shape of the object to be cloaked.

Perhaps an analogy with water waves can be made \cite{Farhat:2008:BCA}.
Our objective is to use the cloaking devices to create an area of still
water, near but outside the cloaking devices, without disturbing the
pattern of waves a certain distance away. Then a boat can be placed in
the area of still water, without disturbing the surrounding waves: the
boat is cloaked, and so are the cloaking devices. The area of still
water is created by destructive interference between the surrounding
waves and anomalously localized waves created by the cloaking devices.
Our cloaking approach has the disadvantage that one needs to know in
advance the incoming probing waves.

For simplicity our analysis is restricted to the two dimensional case,
corresponding to transverse electric or magnetic waves, so the governing
equation is the Helmholtz equation. To begin with we study the
two-dimensional quasistatic problem since the analysis can be carried
further in that case and yields valuable insights. For the Helmholtz equation our
results are purely numerical but provide convincing evidence that
broadband exterior cloaking is possible. 

For the two dimensional quasistatic problem we assume that the
dielectric constant of the background media is constant, i.e., the
voltage is harmonic. Also by $B_r(\Gx)\subset \RR^2$ we will
denote the disk with radius $r$ centered at $\Bx=(\Gx,0)$, and by $f$ we
denote the potential due to exterior sources, that would exist in the
absence of both the cloaking device and object to be cloaked.  The
exterior cloak we propose for the quasistatic problem, consists of one
active cloaking device, which is a simply connected region containing
the origin, along the boundary of which the potential can be prescribed,
and a region to be cloaked $B_\Ga(\Gd)$ with $\Gd>\Ga>0$, that is {\em exterior} to
the cloaking device.  The cloaking is achieved as
follows: according to the assumed known potential $f$ (which could be
obtained by suitably placing probes in the surrounding
medium) the active device generates appropriate fields  such that one
will have very small fields inside $B_\Ga(\Gd)$ and at the same time a very
small total scattering effect, (due to the device and the object),
outside a sufficiently large disk, i.e., $B_\Gg(0)$, with $\Gg>\Ga+\Gd$.
Therefore, an exterior cloak, will, regardless of the probing field,
achieve approximate invisibility of both the active device and any
passive object placed in the disk $B_\Ga(\Gd)$.

The question is then to provide a constructive way to find the potential
at the device in order to achieve cloaking.  First, notice that by
subtracting $f$ from the potential everywhere and applying an inversion
transformation to the problem, namely $z\equiv 1/s$ where $s=x_1+ix_2$,
the question now is to
\begin{multline}
\text{Find a harmonic}~v:\RR^2\rightarrow \RR
 ~\text{such that}~\\
 v \approx 0 ~\text{in}~ B_{1/\Gg}(0) ~\text{and}~ v  \approx  -f
 ~\text{in}~ B_{\Ga_*}({\Gd_*}). 
 \label{1}
\end{multline}
Here we have $\Ga_*=\Ga/(\Gd^2-\Ga^2)$ and $\Gd_*=\Gd/(\Gd^2-\Ga^2)$. 
In fact we only need $v$
to be harmonic outside the image of the cloaking device, but requiring
it to be harmonic in all $\RR^2$ simplifies the problem. If such a
function $v$ exists then its Dirichlet data on the boundary of the
cloaking device gives us the necessary potential one needs to
generate at the surface of the device in order to achieve approximate
invisibility.

By introducing harmonic conjugate potentials one obtains the analytic
extensions, $V$ and $F$, of $v$ and $f$ respectively. Then problem
\eq{1} is equivalent to,
\begin{multline}
\text{Find}~
V:\CC \rightarrow \CC ~\text{analytic, such that}~\\ V \approx 0
~\text{in}~ B_{1 / \Gg}(0) ~\text{and}~ V  \approx  -F ~\text{in}~
B_{\Ga_*}(\Gd_*).
\label{2}
\end{multline}

Since the product of two analytic functions is again analytic, the problem
\eq{2} 
can be equivalently formulated as 
\begin{multline}
\text{Find}~
W:\CC \rightarrow \CC ~\text{analytic, such that}~\\
W \approx 0 ~\text{in}~ B_{1/\Gg}(0) ~\text{and}~ W  \approx  1
~\text{in}~ B_{\Ga_*}(\Gd_*).
\label{3}
\end{multline}
To recover $V$ one needs to multiply $W$ by a polynomial which
approximates $-F$ in $B_{\Ga_*}(\Gd_*)$.

 Next we consider the Hermite interpolation polynomial \mbox{$h:\CC
 \rightarrow \CC$} of
 degree $2n-1$ defined by 
\beqa && h(0)=1,\quad h(\Gd_*)=0,\nonum && 
h^{(j)}(0)=h^{(j)}(\Gd_*)=0\quad\text{for $j = 1,\ldots,n-1$}.
\eeqa{4}
From \eq{4}, by algebraic and combinatoric manipulations together with
an induction argument, it can be shown that
 \beq
 \begin{aligned}
 h(z)&=(z-\Gd_*)^n \sum_{j=0}^{n-1}{z^j\over j!}{d^j\over dy^j}
 \left.\Mp{\frac{1}{(y-\Gd_*)^n}}\right|_{y=0}\\
 &=\Mp{1-{z\over \Gd_*}}^n \sum_{j=0}^{n-1}\Mp{z \over
 \Gd_*}^j\binom{n+j-1}{j}\\
 &={1\over 2} + \Mp{{1\over 2}-{z\over \Gd_*}}
 \sum_{k=0}^{n-1}\Mp{1-{z\over \Gd_*}}^k\Mp{z\over \Gd_*}^k \binom{2k}{k}.
 \end{aligned}
\eeq{hermite} 
Notice that \eq{hermite} implies the symmetry property $h(\Gd_*-z)+h(z)=1$.

\begin{figure}
\begin{center}
\includegraphics[width=0.45\textwidth]{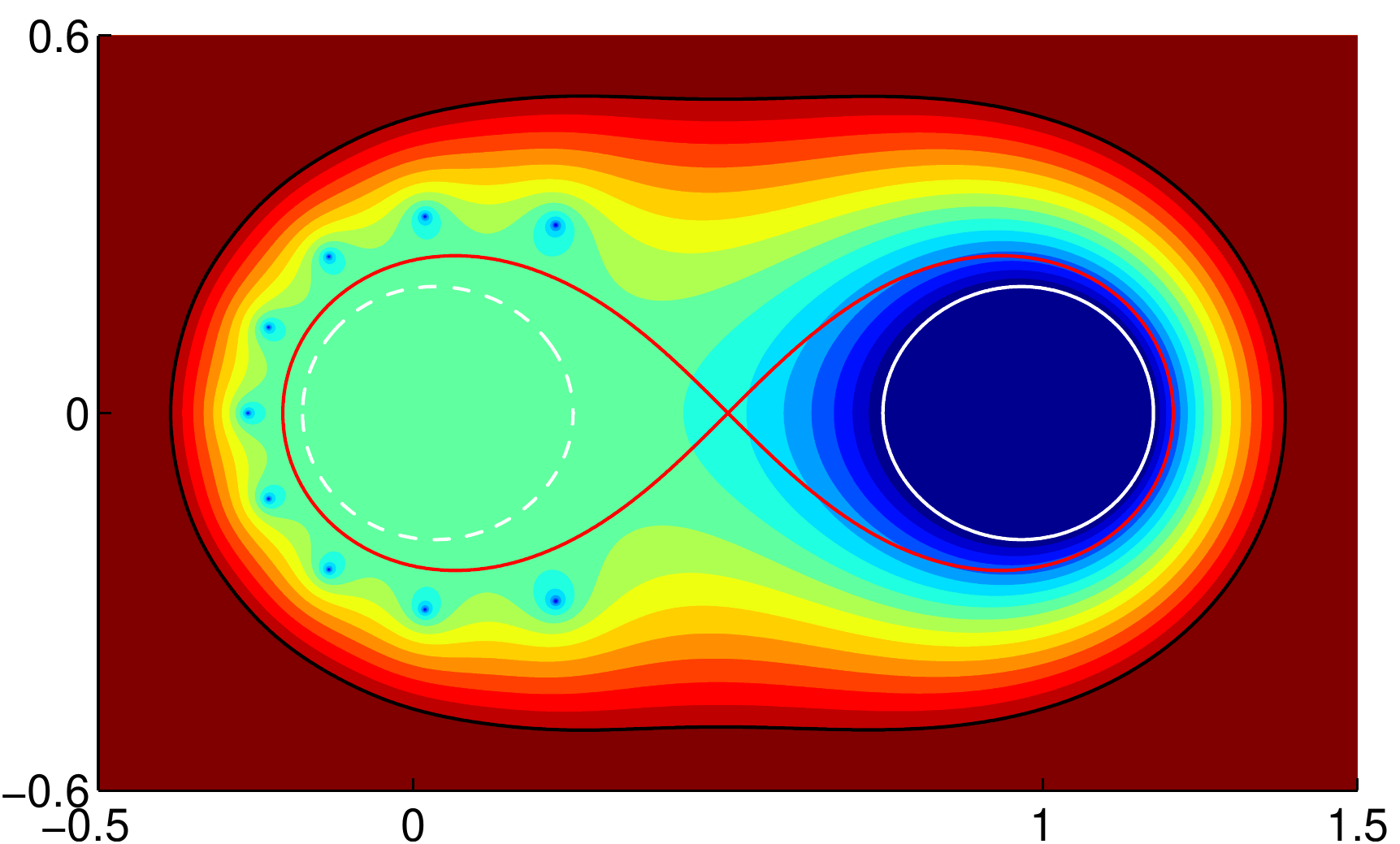}
\end{center}
\caption{Magnitude of the Hermite polynomial $h(z)$ for $\Gd_*=1$ and
degree $n=10$. The color scale is logarithmic from $0.01$ (dark blue)
to $100$ (dark red).}
\label{fig:hermite}
\end{figure}

Our claim that $W(z)=1-h(z)$ satisfies the 
properties in \eq{3} when
$1/\Gg$ and $\Ga_*$ are small enough is strongly supported by
\figref{hermite}: the solid white curve corresponds to the contour
$|h(z)|= 0.01$ and the dashed white curve to $|h(z) - 1| = 0.01$. Also,
the last equality in \eq{hermite} implies that $h(z)$ is in fact a power
series in the variable $z/\Gd_*$, which by the ratio test converges as
$n\to\infty$ in the entire figure eight shaped region
$|z^2-\Gd_* z|<\Gd_*^2/4$ (red curve in \figref{hermite}), and diverges
everywhere outside this region, excluding the boundary. The convergence is uniform in any simply
connected domain that lies strictly within the figure eight.  Therefore
the limit function is analytic in each half of the figure eight, and
from the Taylor series of the limit function at the points $z=0$ and
$z=\Gd_*$, implied by \eq{4}, we deduce that the limit function is one in
the left side of the figure eight, and zero in the right side of the
figure eight.  
To ensure that \eq{3} is satisfied 
we require that the
disks $B_{1/\Gg}(0)$ and $B_{\Ga_*}(\Gd_*)$ lie, respectively, within the
left and right sides of the figure eight, which is the case if both
$1/\Gg$ and $\Ga_*$ are less than $\Gd_*/(2+2\sqrt{2})$.  Notice that one gets
cloaking (in the limit $n\to\infty$) not just within the disk
$B_{\Ga_*}({\Gd_*})$ but in the whole right half of the figure eight.
Thus the scheme based on the Hermite polynomial 
defined in \eq{hermite}
achieves exterior cloaking in the original $s=1/z$ plane for
objects which are sufficiently small compared with $\Gg$. For example with
$\Gd_*=1$, we have $\Ga_*<0.2$, $\Gg>4.9$ and $\Ga=\Ga_*/(\Gd_*^2-\Ga_*^2)<0.2$.

\begin{figure*}
\begin{center}
\begin{tabular}{cc}
 \includegraphics[width=0.34\textwidth]{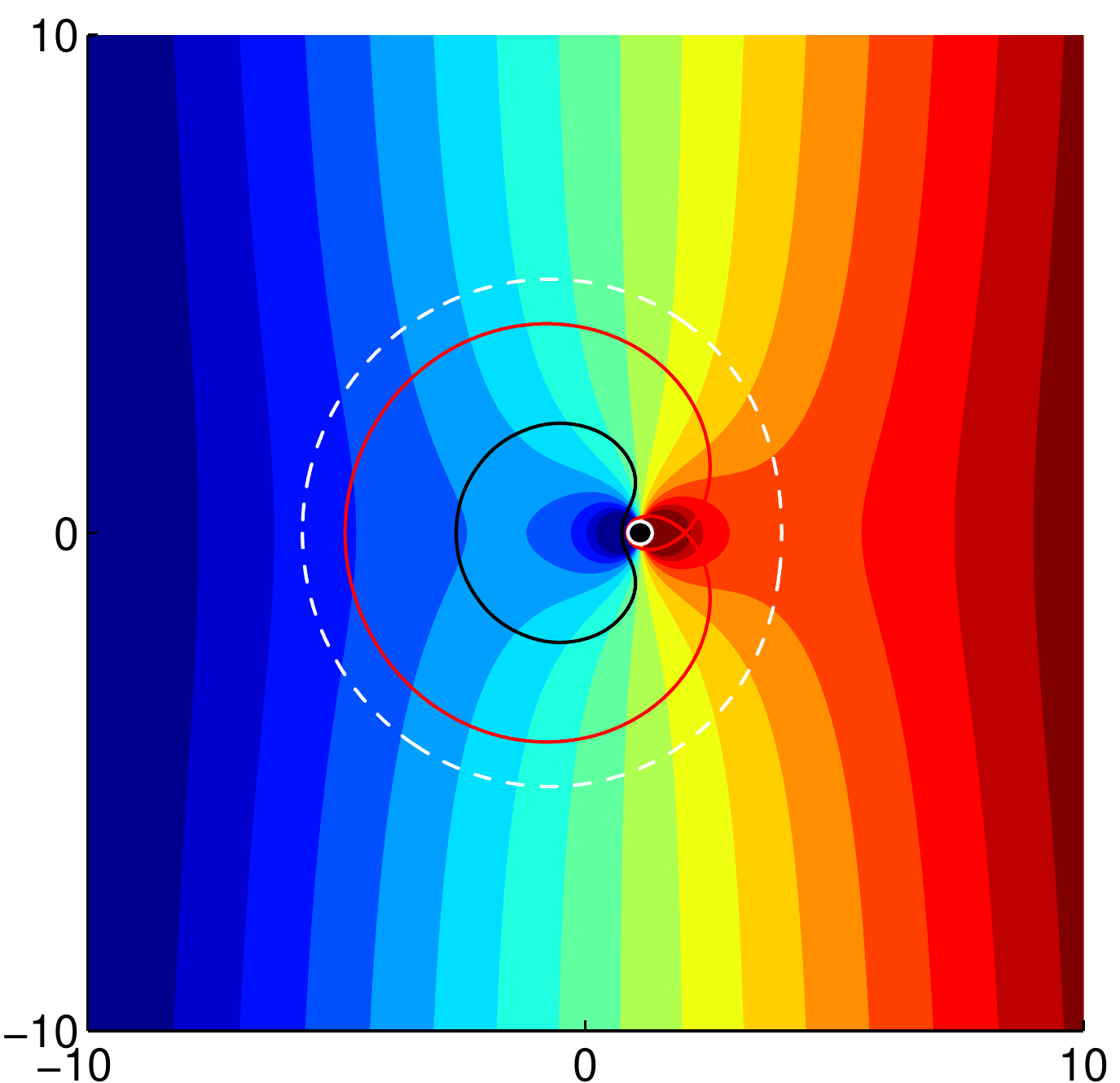} &
 \includegraphics[width=0.34\textwidth]{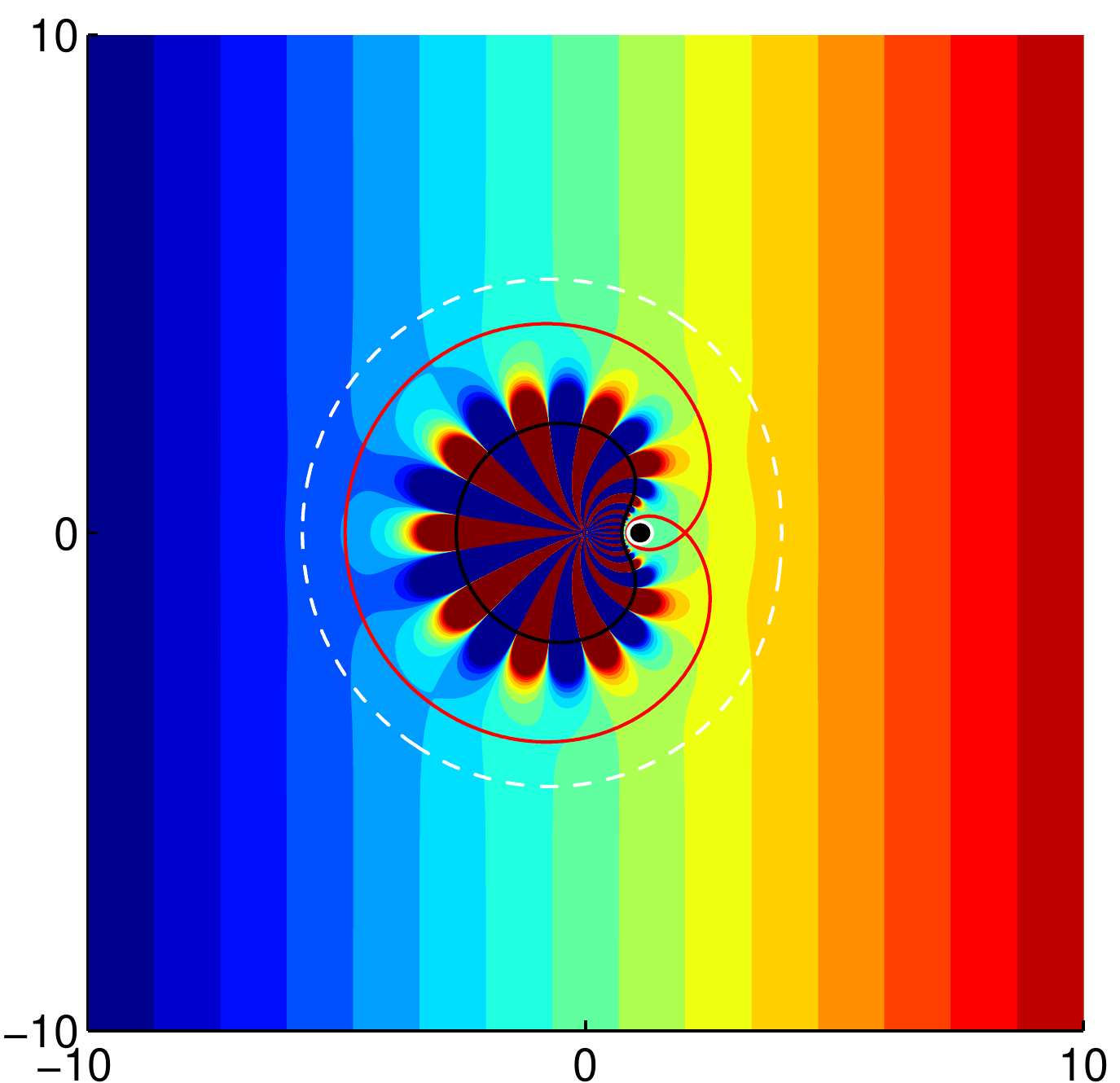}\\
 (a) & (b)
\end{tabular}
\end{center}
\caption{Real part of the total field with the quasistatic cloaking
device (a) inactive and (b) active in the $s$ plane.  The color scale is
linear from $-10$ (dark blue) to $10$ (dark red) and the scatterer
appears in black.}
\label{fig:static:cloak}
\end{figure*}

The cloaking device can be taken to be any simply connected region which
contains the origin, but not the point $(\Gd,0)$. If it is a small
disk, say $B_r(0)$ with $r \ll 1$ then one needs to generate
enormously large potentials at its boundary.  Alternatively its boundary
could be taken as the contour where, say $|h(1/s)|=100$ (black curve in
\figref{hermite} and \fig{static:cloak}), then the cloaking device will
tend to surround the cloaking region as $n\to\infty$.  Such a cloaking
device is demonstrated in \figref{static:cloak} where we wish to cloak a
(almost resonant) disk of radius $0.2$ centered at $s=1.1$ with dielectric constant
$\Ge=-0.99$ and located inside the cloaked region (solid white curve
where $|h(1/s)| = 0.01$). This ``scatterer'' deforms the vertical
equipotential lines of the ``incident'' field $f(s)=s$ when the device
is inactive (\figref{static:cloak}a). When the device is active
the effect of the scatterer is greatly diminished, making it for all
practical purposes invisible: outside the dashed white curve where
$|h(1/s)-1| = 0.01$, the equipotential lines are vertical
(\figref{static:cloak}b). Indeed the discrepancy between the incident
field and the field on the dashed white circle in \figref{static:cloak}b
is of about $1.1\%$ of the incident field or $2.9\%$ of the uncloaked
scattered field, measured in the $L^2$ norm. Due to their association 
with high order multipoles these errors decay very rapidly as $\Gg$
increases. 


One possible extension of the quasistatic cloaking is to the Helmholtz
equation $\Delta u + k^2 u =0$ where $k = 2\pi/\la$ is the wavenumber
and $\la= 2 \pi c_0/\om$ is the wavelength at frequency $\om$ and
constant propagation speed $c_0$. Numerical
simulations suggest it is necessary to take $D\geq 3$  devices located
at points $\Bx_1,\Bx_2,\ldots,\Bx_D$ surrounding the cloaked region to
design good cloaks regardless of the incident field direction. The
cloaked region is for simplicity the disk $|\Bx|\leq \Ga$ and the
devices lie on the circle $|\Bx| = \Gd$. To ensure that objects inside
the cloaked region are hard to observe at locations $|\Bx|\geq \Gg$, the
devices need to create a combined field $u_d$ such that 
\beq
 u_d \approx - u_i ~\text{for}~ |\Bx|\leq \Ga, 
 ~\text{and}~
 u_d \approx 0 ~\text{for}~ |\Bx|\geq \Gg,
\eeq{clkcdt}
where $u_i$ is the field due to exterior sources which would exist in the absence
of the cloaking devices and the object to be cloaked. 
Since we want $u_d$ to be a solution to the Helmholtz equation and decay far
from the devices, we use the following ansatz,
\beq
 u_d(\Bx) = \sum_{m=1}^D \sum_{n=-N}^N b_{m,n} H^{(1)}_n (k|\Bx -
 \Bx_m|) \exp[i n \theta_m ],
\eeq{ud}
where $H^{(1)}_n$ is the $n-$th Hankel function of the first kind and
$\theta_m \equiv \arg(\Bx - \Bx_m)$ is the angle between the vectors
$\Bx-\Bx_m$ and $(1,0)$.  The
coefficients $b_{m,n} \in \CC$ are found numerically by enforcing
\eq{clkcdt} on $N_\Ga$ points $\Bp^\Ga_1, \ldots, \Bp^\Ga_{N_\Ga}$ of the circle
$|\Bx| = \Ga$ and $N_\Gg$ points $\Bp^\Gg_1, \ldots, \Bp^\Gg_{N_\Gg}$ of the
circle $|\Bx| = \Gg$. The resulting linear equations are $\BA \Bb \approx - \Bu_i$
and $\BB \Bb \approx \Bzero$,
where $\Bb \in \CC^{DM}$, $M=2N+1$, is a vector with the coefficients
$b_{m,n}$, and the matrices $\BA \in \CC^{N_\Ga \times DM}$ and $\BB \in
\CC^{N_\Gg \times DM}$ are constructed so that $(\BA \Bb)_j =
u_d(\Bp^\Ga_j)$ and $(\BB \Bb)_j = u_d(\Bp^\Gg_j)$.

Coefficients $\Bb$ satisfying these equations
in the least
squares sense can be obtained via the Singular Value Decomposition (SVD)
in two steps.  First a solution $\Bb_0$ with $\BA \Bb_0 \approx - \Bu_i$
is calculated using the truncated SVD. Then a correction $\Bz$ is found
as a minimizer of $\|\BB (\Bb_0 + \Bz)\|^2$ such that $\BA\Bz=\Bzero$.
The latter linear least squares problem can be easily solved with the
truncated SVD if a basis of the nullspace of $\BA$ is available, which
is the case since the SVD of $\BA$ was computed in the first step. Thus
the coefficients to drive the cloaking devices are $\Bb = \Bb_0 + \Bz$. 

\begin{figure*}
 \begin{center}
 \begin{tabular}{cc}
  \includegraphics[width=0.34\textwidth]{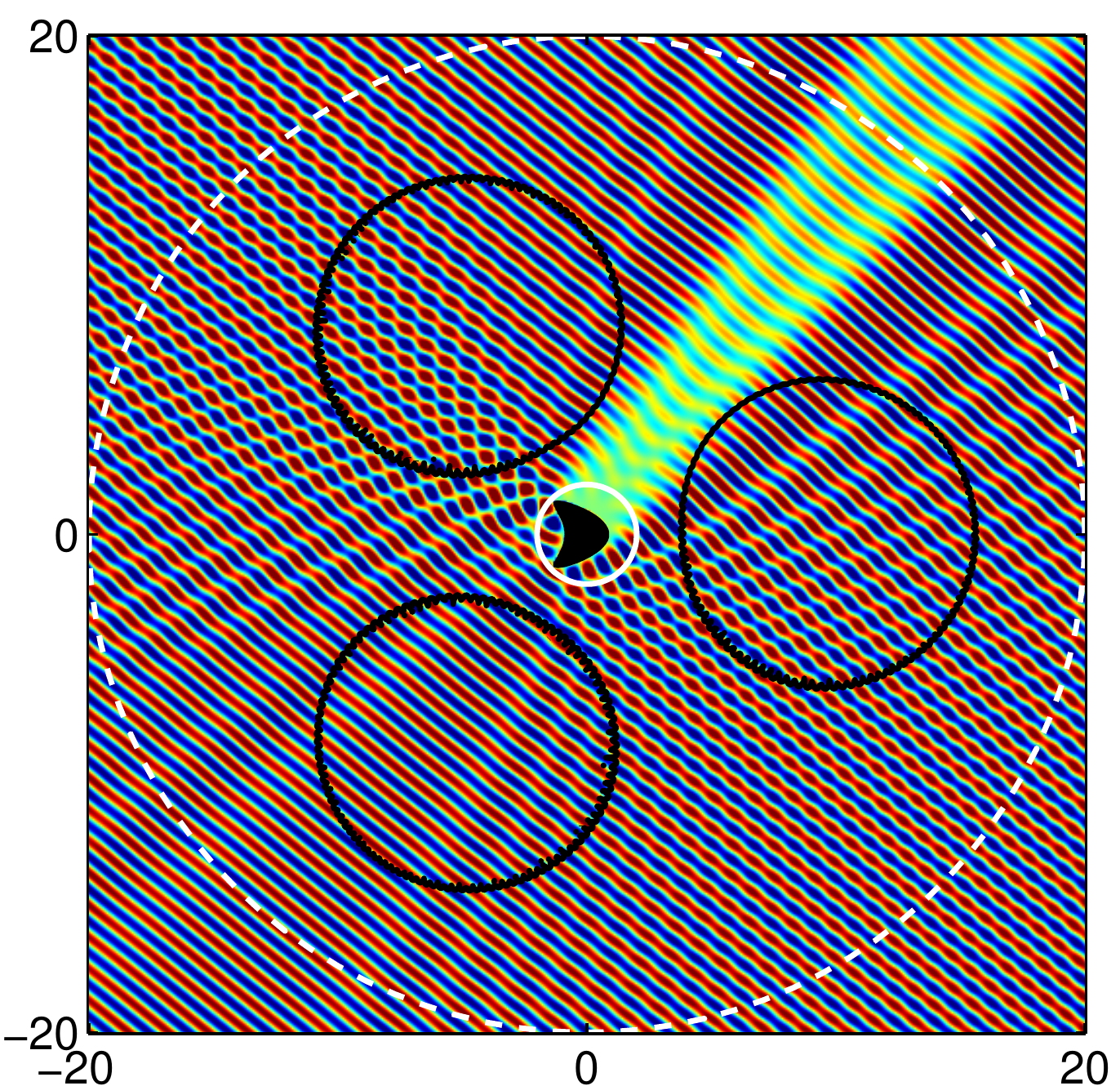} &
  \includegraphics[width=0.34\textwidth]{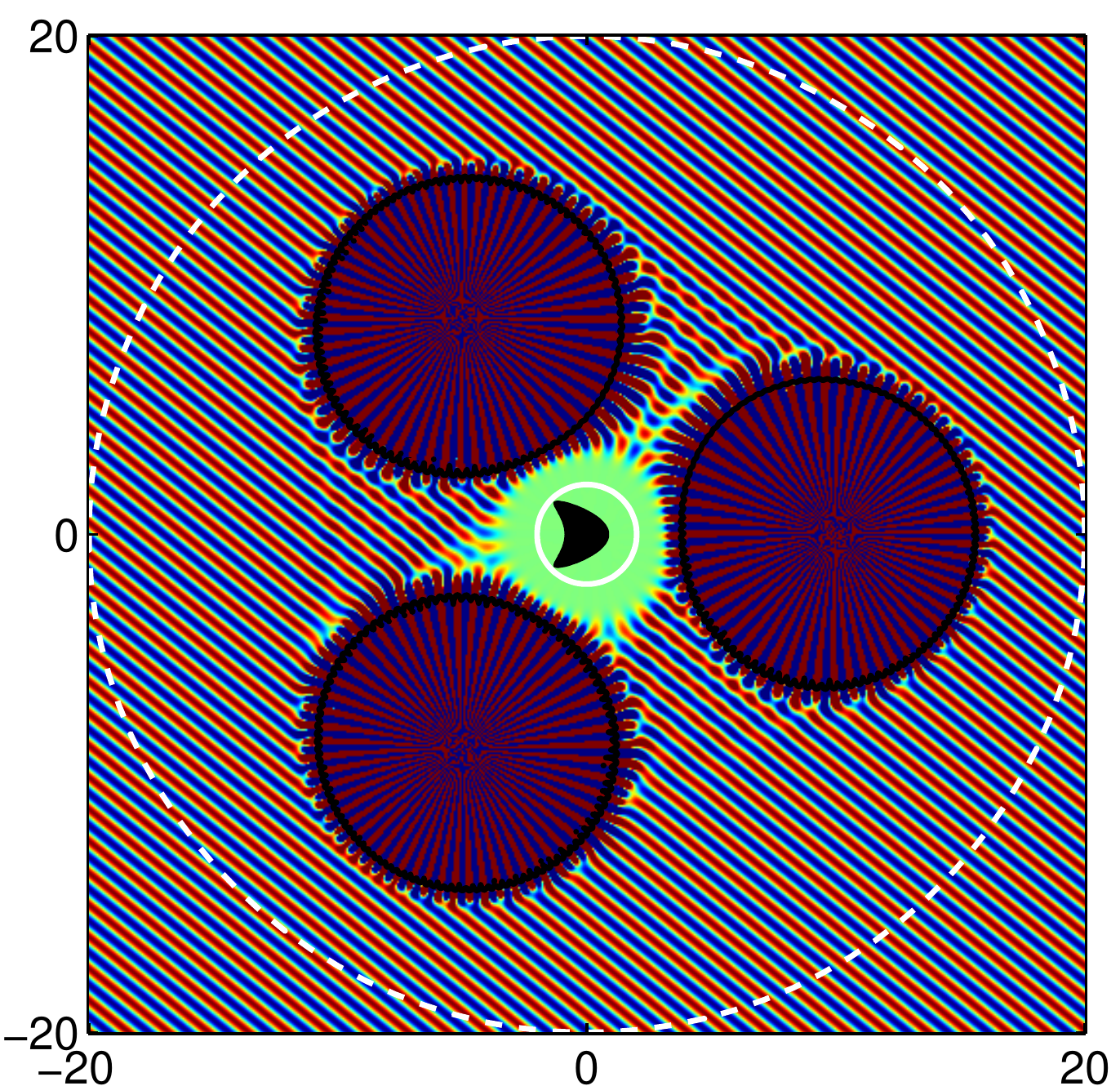}\\
  (a) & (b)
 \end{tabular}
 \end{center}
\caption{Real part of the total field with the Helmholtz cloaking
devices (a) inactive and (b) active. The color scale is linear from -1
(dark blue) to 1 (dark red) and the axis units are $\la$.}
\label{fig:helm}
\end{figure*}

We illustrate this procedure in \figref{helm} with three devices
located $\Gd=10\la$ away from the origin. Here $k=c_0=1$, $\la =
2\pi$ and we apply a plane wave incident field $u_i(\Bx) = \exp[ik\Bx
\cdot \Bd]$ with $\Bd = (\cos(2\pi/7), \sin(2\pi/7))$. The cloaked
region is the solid white circle with radius $\Ga=2\la$.  Invisibility
is enforced on the dashed white circle of radius $\Gg=20\la$. The
control points $\Bp_j^\Gg$ and $\Bp_j^\Ga$ are uniformly spaced and less
than $\la/2$ apart on their respective circles and $N=57$ terms were
used for the ansatz \eq{ud}. Finally the scattered field resulting from
the impenetrable ``kite'' obstacle with sound-soft (homogeneous Dirichlet)
boundary conditions is computed using the boundary integral equation
approach in \cite{Colton:1998:IAE}.

As shown in \figref{helm}b. when the cloaking devices are active they
create a ``quiet'' region where the wave field is close to zero. An
object lying in this region is practically invisible because both the
scattered and device's fields are for all practical purposes
undetectable outside of the dashed white circle. The field outside the
dashed white circle is nearly identical to the incident plane wave: the
discrepancy on that circle is of about $1.5\times 10^{-5}\%$ of the
incident field or $6.8 \times 10^{-5}\%$ of the uncloaked scattered field,
measured with the $L^2$ norm. Of course these relative errors depend on
the particular choice of the cutoffs for the singular values in our
two step approach.

By Green's identities, these point devices could be replaced by a
bounded region containing $\Bx_1,\ldots,\Bx_D$ but not the cloaked
region, provided the region's boundary has a controllable single- and
double-layer potential (i.e. a point source and dipole density
\cite{Colton:1998:IAE}). One example is to take $D$ disjoint disks,
however if their radii are too small then the strength of the potentials
on the disk boundaries would be enormous (since for $n\neq 0$,
$H^{(1)}_n(r) = \CO(r^{-|n|})$ as $r\to 0$). A natural question is
whether one can get exterior cloaking with reasonable field magnitudes.
The  black curves representing the contours $|u_d(\Bx)|=100$ in
\figref{helm} suggest this is possible because they resemble three
disjoint disks.

A complete mathematical discussion, including the rigorous analysis for
the arguments for the quasistatic case, together with a different
analytical approach, will be included in a forthcoming publication. The
theory for the Helmholtz problem remains an object of current research.
We anticipate that the results extend to three-dimensions and to
the full Maxwell equations but this remains to be explored.

\begin{acknowledgments}
The authors are grateful for support from the National Science
Foundation through grant DMS-070978. 
\end{acknowledgments}

\bibliographystyle{apsrev}
\bibliography{/u/ma/milton/tcbook,/u/ma/milton/newref}

\ifx \bblindex \undefined \def \bblindex #1{} \fi\ifx \bblindex \undefined \def
  \bblindex #1{} \fi
\begin{thebibliography}{30}
\expandafter\ifx\csname natexlab\endcsname\relax\def\natexlab#1{#1}\fi
\expandafter\ifx\csname bibnamefont\endcsname\relax
  \def\bibnamefont#1{#1}\fi
\expandafter\ifx\csname bibfnamefont\endcsname\relax
  \def\bibfnamefont#1{#1}\fi
\expandafter\ifx\csname citenamefont\endcsname\relax
  \def\citenamefont#1{#1}\fi
\expandafter\ifx\csname url\endcsname\relax
  \def\url#1{\texttt{#1}}\fi
\expandafter\ifx\csname urlprefix\endcsname\relax\def\urlprefix{URL }\fi
\providecommand{\bibinfo}[2]{#2}
\providecommand{\eprint}[2][]{\url{#2}}

\bibitem[{\citenamefont{Dolin}(1961)}]{Dolin:1961:PCT}
\bibinfo{author}{\bibfnamefont{L.~S.} \bibnamefont{Dolin}},
  \bibinfo{journal}{Izv. Vyssh. Uchebn. Zaved. Radiofizika}
  \textbf{\bibinfo{volume}{4}}, \bibinfo{pages}{964} (\bibinfo{year}{1961}).

\bibitem[{\citenamefont{Kerker}(1975)}]{Kerker:1975:IB}
\bibinfo{author}{\bibfnamefont{M.}~\bibnamefont{Kerker}},
  \bibinfo{journal}{J. Opt. Soc. Am.}
  \textbf{\bibinfo{volume}{65}}, \bibinfo{pages}{376} (\bibinfo{year}{1975}).

\bibitem[{\citenamefont{Al\'u and Engheta}(2008)}]{Alu:2008:PMC}
\bibinfo{author}{\bibfnamefont{A.}~\bibnamefont{Al\'u}} \bibnamefont{and}
  \bibinfo{author}{\bibfnamefont{N.}~\bibnamefont{Engheta}},
  \bibinfo{journal}{J. Opt. A: Pure Appl. Opt.} \textbf{\bibinfo{volume}{10}},
  \bibinfo{pages}{093002} (\bibinfo{year}{2008}).

\bibitem[{\citenamefont{Greenleaf et~al.}(2009)\citenamefont{Greenleaf,
  Kurylev, Lassas, and Uhlmann}}]{Greenleaf:2009:CDE}
\bibinfo{author}{\bibfnamefont{A.}~\bibnamefont{Greenleaf}},
  \bibinfo{author}{\bibfnamefont{Y.}~\bibnamefont{Kurylev}},
  \bibinfo{author}{\bibfnamefont{M.}~\bibnamefont{Lassas}}, \bibnamefont{and}
  \bibinfo{author}{\bibfnamefont{G.}~\bibnamefont{Uhlmann}},
  \bibinfo{journal}{SIAM Rev.} \textbf{\bibinfo{volume}{51}},
  \bibinfo{pages}{3} (\bibinfo{year}{2009}).

\bibitem[{\citenamefont{Al\'u and Engheta}(2005)}]{Alu:2005:ATP}
\bibinfo{author}{\bibfnamefont{A.}~\bibnamefont{Al\'u}} \bibnamefont{and}
  \bibinfo{author}{\bibfnamefont{N.}~\bibnamefont{Engheta}},
  \bibinfo{journal}{Phys. Rev. E} \textbf{\bibinfo{volume}{72}},
  \bibinfo{pages}{016623} (\bibinfo{year}{2005}).

\bibitem[{\citenamefont{Greenleaf et~al.}(2003)\citenamefont{Greenleaf, Lassas,
  and Uhlmann}}]{Greenleaf:2003:ACC}
\bibinfo{author}{\bibfnamefont{A.}~\bibnamefont{Greenleaf}},
  \bibinfo{author}{\bibfnamefont{M.}~\bibnamefont{Lassas}}, \bibnamefont{and}
  \bibinfo{author}{\bibfnamefont{G.}~\bibnamefont{Uhlmann}},
  \bibinfo{journal}{Physiol. Meas.} \textbf{\bibinfo{volume}{24}},
  \bibinfo{pages}{413} (\bibinfo{year}{2003}).

\bibitem[{\citenamefont{Pendry et~al.}(2006)\citenamefont{Pendry, Schurig, and
  Smith}}]{Pendry:2006:CEM}
\bibinfo{author}{\bibfnamefont{J.~B.} \bibnamefont{Pendry}},
  \bibinfo{author}{\bibfnamefont{D.}~\bibnamefont{Schurig}}, \bibnamefont{and}
  \bibinfo{author}{\bibfnamefont{D.~R.} \bibnamefont{Smith}},
  \bibinfo{journal}{Science} \textbf{\bibinfo{volume}{312}},
  \bibinfo{pages}{1780} (\bibinfo{year}{2006}).

\bibitem[{\citenamefont{Leonhardt}(2006)}]{Leonhardt:2006:OCM}
\bibinfo{author}{\bibfnamefont{U.}~\bibnamefont{Leonhardt}},
  \bibinfo{journal}{Science} \textbf{\bibinfo{volume}{312}},
  \bibinfo{pages}{1777} (\bibinfo{year}{2006}).

\bibitem[{\citenamefont{Li and Pendry}(2008)}]{Li:2008:HUC}
\bibinfo{author}{\bibfnamefont{J.}~\bibnamefont{Li}} \bibnamefont{and}
  \bibinfo{author}{\bibfnamefont{J.~B.} \bibnamefont{Pendry}},
  \bibinfo{journal}{Phys. Rev. Lett.} \textbf{\bibinfo{volume}{101}},
  \bibinfo{pages}{203901} (\bibinfo{year}{2008}).

\bibitem[{\citenamefont{Miller}(2006)}]{Miller:2007:PC}
\bibinfo{author}{\bibfnamefont{D.~A.~B.} \bibnamefont{Miller}},
  \bibinfo{journal}{Opt. Express} \textbf{\bibinfo{volume}{14}},
  \bibinfo{pages}{12457} (\bibinfo{year}{2006}).

\bibitem[{\citenamefont{Liu et~al.}(2009)\citenamefont{Liu, Ji, Mock, Chin,
  Cui, and Smith}}]{Liu:2009:BGP}
\bibinfo{author}{\bibfnamefont{R.}~\bibnamefont{Liu}},
  \bibinfo{author}{\bibfnamefont{C.}~\bibnamefont{Ji}},
  \bibinfo{author}{\bibfnamefont{J.~J.} \bibnamefont{Mock}},
  \bibinfo{author}{\bibfnamefont{J.~Y.} \bibnamefont{Chin}},
  \bibinfo{author}{\bibfnamefont{T.~J.} \bibnamefont{Cui}}, \bibnamefont{and}
  \bibinfo{author}{\bibfnamefont{D.~R.} \bibnamefont{Smith}},
  \bibinfo{journal}{Science} \textbf{\bibinfo{volume}{323}},
  \bibinfo{pages}{366} (\bibinfo{year}{2009}).

\bibitem[{\citenamefont{Valentine et~al.}(2009)\citenamefont{Valentine, Li,
  Zentgraf, Bartal, and Zhang}}]{Valentine:2009:OCD}
\bibinfo{author}{\bibfnamefont{J.}~\bibnamefont{Valentine}},
  \bibinfo{author}{\bibfnamefont{J.}~\bibnamefont{Li}},
  \bibinfo{author}{\bibfnamefont{T.}~\bibnamefont{Zentgraf}},
  \bibinfo{author}{\bibfnamefont{G.}~\bibnamefont{Bartal}}, \bibnamefont{and}
  \bibinfo{author}{\bibfnamefont{X.}~\bibnamefont{Zhang}},
  \bibinfo{journal}{Nat. Mater.}  (\bibinfo{year}{2009}),
  \bibinfo{note}{published online doi:10.1038/nmat2461}.

\bibitem[{\citenamefont{Gabrielli et~al.}(2009)\citenamefont{Gabrielli,
  Cardenas, Poitras, and Lipson}}]{Gabrielli:2009:COF}
\bibinfo{author}{\bibfnamefont{L.~H.} \bibnamefont{Gabrielli}},
  \bibinfo{author}{\bibfnamefont{J.}~\bibnamefont{Cardenas}},
  \bibinfo{author}{\bibfnamefont{C.~B.} \bibnamefont{Poitras}},
  \bibnamefont{and} \bibinfo{author}{\bibfnamefont{M.}~\bibnamefont{Lipson}}
  (\bibinfo{year}{2009}), \bibinfo{note}{arXiv:0904.3508v1 [physics.optics]}.

\bibitem[{\citenamefont{Farhat et~al.}(2008)\citenamefont{Farhat, Enoch,
  Guenneau, and Movchan}}]{Farhat:2008:BCA}
\bibinfo{author}{\bibfnamefont{M.}~\bibnamefont{Farhat}},
  \bibinfo{author}{\bibfnamefont{S.}~\bibnamefont{Enoch}},
  \bibinfo{author}{\bibfnamefont{S.}~\bibnamefont{Guenneau}}, \bibnamefont{and}
  \bibinfo{author}{\bibfnamefont{A.~B.} \bibnamefont{Movchan}},
  \bibinfo{journal}{Phys. Rev. Lett.} \textbf{\bibinfo{volume}{101}},
  \bibinfo{pages}{134501} (\bibinfo{year}{2008}).

\bibitem[{\citenamefont{Schurig et~al.}(2006)\citenamefont{Schurig, Mock,
  Justice, Cummer, Pendry, Starr, and Smith}}]{Schurig:2006:MEC}
\bibinfo{author}{\bibfnamefont{D.}~\bibnamefont{Schurig}},
  \bibinfo{author}{\bibfnamefont{J.~J.} \bibnamefont{Mock}},
  \bibinfo{author}{\bibfnamefont{B.~J.} \bibnamefont{Justice}},
  \bibinfo{author}{\bibfnamefont{S.~A.} \bibnamefont{Cummer}},
  \bibinfo{author}{\bibfnamefont{J.~B.} \bibnamefont{Pendry}},
  \bibinfo{author}{\bibfnamefont{A.~F.} \bibnamefont{Starr}}, \bibnamefont{and}
  \bibinfo{author}{\bibfnamefont{D.~R.} \bibnamefont{Smith}},
  \bibinfo{journal}{Science} \textbf{\bibinfo{volume}{314}},
  \bibinfo{pages}{977} (\bibinfo{year}{2006}).

\bibitem[{\citenamefont{Greenleaf et~al.}(2007)\citenamefont{Greenleaf,
  Kurylev, Lassas, and Uhlmann}}]{Greenleaf:2007:FWI}
\bibinfo{author}{\bibfnamefont{A.}~\bibnamefont{Greenleaf}},
  \bibinfo{author}{\bibfnamefont{Y.}~\bibnamefont{Kurylev}},
  \bibinfo{author}{\bibfnamefont{M.}~\bibnamefont{Lassas}}, \bibnamefont{and}
  \bibinfo{author}{\bibfnamefont{G.}~\bibnamefont{Uhlmann}},
  \bibinfo{journal}{Commun. Math. Phys.}
  \textbf{\bibinfo{volume}{275}}, \bibinfo{pages}{749}
  (\bibinfo{year}{2007})
  .

\bibitem[{\citenamefont{Kohn et~al.}(2008{\natexlab{a}})\citenamefont{Kohn,
  Shen, Vogelius, and Weinstein}}]{Kohn:2008:CCV}
\bibinfo{author}{\bibfnamefont{R.~V.} \bibnamefont{Kohn}},
  \bibinfo{author}{\bibfnamefont{H.}~\bibnamefont{Shen}},
  \bibinfo{author}{\bibfnamefont{M.~S.} \bibnamefont{Vogelius}},
  \bibnamefont{and} \bibinfo{author}{\bibfnamefont{M.~I.}
  \bibnamefont{Weinstein}}, \bibinfo{journal}{Inverse Prob.}
  \textbf{\bibinfo{volume}{24}}, \bibinfo{pages}{015016}
  (\bibinfo{year}{2008}{\natexlab{a}}).

\bibitem[{\citenamefont{Kohn et~al.}(2008{\natexlab{b}})\citenamefont{Kohn,
  Onofrei, Vogelius, and Weinstein}}]{Kohn:2009:CCV}
\bibinfo{author}{\bibfnamefont{R.~V.} \bibnamefont{Kohn}},
  \bibinfo{author}{\bibfnamefont{D.}~\bibnamefont{Onofrei}},
  \bibinfo{author}{\bibfnamefont{M.~S.} \bibnamefont{Vogelius}},
  \bibnamefont{and} \bibinfo{author}{\bibfnamefont{M.~I.}
  \bibnamefont{Weinstein}} (\bibinfo{year}{2008}{\natexlab{b}}),
  \bibinfo{note}{in preparation}.

\bibitem[{\citenamefont{Cai et~al.}(2007{\natexlab{a}})\citenamefont{Cai,
  Chettiar, Kildishev, and Shalaev}}]{Cai:2007:OCM}
\bibinfo{author}{\bibfnamefont{W.}~\bibnamefont{Cai}},
  \bibinfo{author}{\bibfnamefont{U.~K.} \bibnamefont{Chettiar}},
  \bibinfo{author}{\bibfnamefont{A.~V.} \bibnamefont{Kildishev}},
  \bibnamefont{and} \bibinfo{author}{\bibfnamefont{V.~M.}
  \bibnamefont{Shalaev}}, \bibinfo{journal}{Nat. Photonics}
  \textbf{\bibinfo{volume}{1}}, \bibinfo{pages}{224}
  (\bibinfo{year}{2007}{\natexlab{a}}).

\bibitem[{\citenamefont{Cai et~al.}(2007{\natexlab{b}})\citenamefont{Cai,
  Chettiar, Kildishev, Shalaev, and Milton}}]{Cai:2007:NMC}
\bibinfo{author}{\bibfnamefont{W.}~\bibnamefont{Cai}},
  \bibinfo{author}{\bibfnamefont{U.~K.} \bibnamefont{Chettiar}},
  \bibinfo{author}{\bibfnamefont{A.~V.} \bibnamefont{Kildishev}},
  \bibinfo{author}{\bibfnamefont{V.~M.} \bibnamefont{Shalaev}},
  \bibnamefont{and} \bibinfo{author}{\bibfnamefont{G.~W.}
  \bibnamefont{Milton}}, \bibinfo{journal}{Appl. Phys. Lett.}
  \textbf{\bibinfo{volume}{91}}, \bibinfo{pages}{111105}
  (\bibinfo{year}{2007}{\natexlab{b}}).

\bibitem[{\citenamefont{Milton and Nicorovici}(2006)}]{Milton:2006:CEA}
\bibinfo{author}{\bibfnamefont{G.~W.} \bibnamefont{Milton}} \bibnamefont{and}
  \bibinfo{author}{\bibfnamefont{N.-A.~P.} \bibnamefont{Nicorovici}},
  \bibinfo{journal}{Proc. R. Soc. A} \textbf{\bibinfo{volume}{462}},
  \bibinfo{pages}{3027} (\bibinfo{year}{2006}).

\bibitem[{\citenamefont{Nicorovici et~al.}(2007)\citenamefont{Nicorovici,
  Milton, McPhedran, and Botten}}]{Nicorovici:2007:OCT}
\bibinfo{author}{\bibfnamefont{N.-A.~P.} \bibnamefont{Nicorovici}},
  \bibinfo{author}{\bibfnamefont{G.~W.} \bibnamefont{Milton}},
  \bibinfo{author}{\bibfnamefont{R.~C.} \bibnamefont{McPhedran}},
  \bibnamefont{and} \bibinfo{author}{\bibfnamefont{L.~C.}
  \bibnamefont{Botten}}, \bibinfo{journal}{Opt. Express}
  \textbf{\bibinfo{volume}{15}}, \bibinfo{pages}{6314} (\bibinfo{year}{2007}).

\bibitem[{\citenamefont{Milton et~al.}(2008)\citenamefont{Milton, Nicorovici,
  McPhedran, Cherednichenko, and Jacob}}]{Milton:2008:SFG}
\bibinfo{author}{\bibfnamefont{G.~W.} \bibnamefont{Milton}},
  \bibinfo{author}{\bibfnamefont{N.-A.~P.} \bibnamefont{Nicorovici}},
  \bibinfo{author}{\bibfnamefont{R.~C.} \bibnamefont{McPhedran}},
  \bibinfo{author}{\bibfnamefont{K.}~\bibnamefont{Cherednichenko}},
  \bibnamefont{and} \bibinfo{author}{\bibfnamefont{Z.}~\bibnamefont{Jacob}},
  \bibinfo{journal}{New J. Phys.} \textbf{\bibinfo{volume}{10}},
  \bibinfo{pages}{115021} (\bibinfo{year}{2008}).

\bibitem[{\citenamefont{Milton et~al.}(2007)\citenamefont{Milton, Nicorovici,
  and McPhedran}}]{Milton:2006:OPL}
\bibinfo{author}{\bibfnamefont{G.~W.} \bibnamefont{Milton}},
  \bibinfo{author}{\bibfnamefont{N.-A.~P.} \bibnamefont{Nicorovici}},
  \bibnamefont{and} \bibinfo{author}{\bibfnamefont{R.~C.}
  \bibnamefont{McPhedran}}, \bibinfo{journal}{Physica B}
  \textbf{\bibinfo{volume}{394}}, \bibinfo{pages}{171} (\bibinfo{year}{2007}).

\bibitem[{\citenamefont{Veselago}(1967)}]{Veselago:1967:ESS}
\bibinfo{author}{\bibfnamefont{V.~G.} \bibnamefont{Veselago}},
  \bibinfo{journal}{Sov. Phys. Usp.},
  \textbf{\bibinfo{volume}{10}},
  \bibinfo{pages}{509--514} (\bibinfo{year}{1968}).


\bibitem[{\citenamefont{Nicorovici et~al.}(1994)\citenamefont{Nicorovici,
  McPhedran, and Milton}}]{Nicorovici:1994:ODP}
\bibinfo{author}{\bibfnamefont{N.~A.} \bibnamefont{Nicorovici}},
  \bibinfo{author}{\bibfnamefont{R.~C.} \bibnamefont{McPhedran}},
  \bibnamefont{and} \bibinfo{author}{\bibfnamefont{G.~W.}
  \bibnamefont{Milton}}, \bibinfo{journal}{Phys. Rev. B}
  \textbf{\bibinfo{volume}{49}}, \bibinfo{pages}{8479}
  (\bibinfo{year}{1994}).

\bibitem[{\citenamefont{Pendry}(2000)}]{Pendry:2000:NRM}
\bibinfo{author}{\bibfnamefont{J.~B.} \bibnamefont{Pendry}},
  \bibinfo{journal}{Phys. Rev. Lett.} \textbf{\bibinfo{volume}{85}},
  \bibinfo{pages}{3966} (\bibinfo{year}{2000}).

\bibitem[{\citenamefont{Bruno and Lintner}(2007)}]{Bruno:2007:SCS}
\bibinfo{author}{\bibfnamefont{O.~P.} \bibnamefont{Bruno}} \bibnamefont{and}
  \bibinfo{author}{\bibfnamefont{S.}~\bibnamefont{Lintner}},
  \bibinfo{journal}{J. Appl. Phys.} \textbf{\bibinfo{volume}{102}},
  \bibinfo{pages}{124502} (\bibinfo{year}{2007}).

\bibitem[{\citenamefont{Lai et~al.}(2009)\citenamefont{Lai, Chen, Zhang, and
  Chan}}]{Lai:2009:CMI}
\bibinfo{author}{\bibfnamefont{Y.}~\bibnamefont{Lai}},
  \bibinfo{author}{\bibfnamefont{H.}~\bibnamefont{Chen}},
  \bibinfo{author}{\bibfnamefont{Z.-Q.} \bibnamefont{Zhang}}, \bibnamefont{and}
  \bibinfo{author}{\bibfnamefont{C.~T.} \bibnamefont{Chan}},
  \bibinfo{journal}{Phys. Rev. Lett.} \textbf{\bibinfo{volume}{102}},
  \bibinfo{pages}{093901} (\bibinfo{year}{2009}).

\bibitem[{\citenamefont{Colton and Kress}(1998)}]{Colton:1998:IAE}
\bibinfo{author}{\bibfnamefont{D.}~\bibnamefont{Colton}} \bibnamefont{and}
  \bibinfo{author}{\bibfnamefont{R.}~\bibnamefont{Kress}},
  \emph{\bibinfo{title}{Inverse acoustic and electromagnetic scattering
  theory}}, vol.~\bibinfo{volume}{93} of \emph{\bibinfo{series}{Applied
  Mathematical Sciences}} (\bibinfo{publisher}{Springer-Verlag},
  \bibinfo{address}{Berlin}, \bibinfo{year}{1998}), \bibinfo{edition}{2nd} ed..

\end{thebibliography}

\end{document}